\documentclass[aps,prl,superscriptaddress,showpacs,twocolumn,footinbib]{revtex4-2}
\usepackage[pdftex]{graphicx}
\usepackage{amsmath}
\usepackage{xspace}
\usepackage{color}
\usepackage[colorlinks=true,linkcolor=blue,urlcolor=blue,citecolor=blue]{hyperref}
\usepackage[normalem]{ulem}
\usepackage{hyperref}
\usepackage{todonotes} 

\usepackage{color}
\definecolor{BV}{rgb}{0.1,0.,0.6}
\definecolor{R}{rgb}{0.9,0,0}
\definecolor{G}{rgb}{0.2,0.8,0.2}
\usepackage{comment}

\begin{document}

\title{Emergence of anyonic correlations from spin and charge dynamics in one dimension}

\author{Oleksandr Gamayun}
\affiliation{London Institute for Mathematical Sciences, Royal Institution, 21 Albermarle St, London W1S 4BS, UK}
\affiliation{Faculty of Physics, University of Warsaw, ul. Pasteura 5, 02-093 Warsaw, Poland}

\author{Eoin Quinn}
\affiliation{Universit\'e Paris-Saclay, CNRS, LPTMS, 91405, Orsay, France}

\author{Kemal Bidzhiev}
\affiliation{PASQAL, 7 rue L\'eonard de Vinci, 91300 Massy, France}

\author{Mikhail B. Zvonarev}
\affiliation{Universit\'e Paris-Saclay, CNRS, LPTMS, 91405, Orsay, France}

\date{\today}

\begin{abstract}

We propose a transformation for spin and charge degrees of freedom in one-dimensional lattice systems, constrained to have no doubly occupied sites, that allows direct access to the dynamical correlations of the system. The transformation delivers particle creation and annihilation operators in a form of a spinless particle and a non-local operator acting on the space of states of a spin-$1/2$ chain. This permits a decomposition of dynamical correlation functions as a convolution of those for impenetrable anyons together with those of a spin chain. Further analysis can be done by methods tailored for each part of the convolution, greatly increasing the impact and flexibility of the approach.



\end{abstract}

\maketitle

The physics of many-body quantum systems incorporates effects from interaction and statistics of bare particles. The emerging quasi-particles could inherit the statistics of their non-interacting peers, free fermions turning into a Fermi liquid, and free bosons into a Bose-Einstein condensate. Reducing a system's dimensionality enhances interaction effects and masks out signatures of the statistics of the constitutent particles. In one dimension, arbitrarily weak repulsion precludes a macroscopic occupation of a single state with the zero momentum, that is, destroys the Bose-Einstein condensate~\cite{pitaevskii_book_BEC}. Furthermore, interactions may transform bosonic excitation spectrum into a fermionic one, an example being the bosons repelling each other through a $\delta$-function potential of infinite strength, the system known as the Tonks-Girardeau gas, whose excitation spectrum is identical to that of a free Fermi gas~\cite{girardeau_impurity_TG_60}. 

The interplay of spin and charge degrees of freedom could be particularly intricate in one dimension. Systems having linear excitation spectrum at low energies fall into a Luttinger liquid (LL) universality class regardless of the statistics of the bare particles. Spin and charge degrees of freedom of the microscopic theory are represented by commuting terms in the LL Hamiltonian and factor out in the dynamical correlation functions, the phenomenon referred to as spin-charge separation~\cite{gogolin_1dbook,giamarchi_book_1d}. Accounting for non-linearities of the excitation spectrum within the effective field theory approach requires proper modification of the LL description, the cases studied recently being spin and charge dynamics above the highly degenerate ground state (spin-incoherent regime, Refs.~\cite{cheianov_spin_decoherent_short_04, fiete_spin_decoherent_04,fiete_SI_07}), in presence of the quadratic branch of the excitation spectrum (itinerant ferromagnetic regime, Refs.~\cite{zvonarev_ferrobosons_07,akhanjee_ferrobosons_07,kamenev_spinor_bosons_09, zvonarev_Yang_Gaudin_09, zvonarev_BoseHubb_09}), and in the vicinity of the edge of excitation spectrum, Ref.~\cite{imambekov_review_12}. Whether and how the concept of the spin-charge separation may be extended beyond the LL effective field theory description is a challenging open question, relevant, in particular, for ultracold gas experiments~\cite{senarathe_spincharge_fermi_22}.

Studying systems with no double occupancy (NDO) constraint (any two particles cannot occupy the same lattice site) is a must for understanding how spin and charge degrees of freedom are coupled at all energy scales. Disregarding the unoccupied sites (``squeezing'' the lattice) reduces the space of states of the original system containing $N$ spin-$1/2$ particles to the space of states of the spin-$1/2$ chain of length $N$. The state of individual spins on the squeezed lattice could be controlled and manipulated directly by ultracold quantum gas microscopy~\cite{hilker_string_17, salomon_incommensurate_19, vijayan_deconf_Hubbard_20}. On the theory side, some dynamical correlation functions have been evaluated by making use of the coordinate representation for the many body wave functions, whose structure is very special due to the NDO constraint~\cite{ogata_BA_90,zabrodin_Fermi_strong_89,zabrodin_single_particle_90, izergin_impenetrable_hubbard_98}. The formalism of the second quantization, expressing basic microscopic fields of the system in terms of the collective spin and charge variables, could serve as a systemic approach revealing contributions from spin and charge dynamics into any correlation function. However, such a formalism has not been developed so far.

In this Letter we present a transformation from the spin-$1/2$ fermions subjected to the NDO constraint to the collective charge (spinless fermions on a lattice) and spin (spin-$1/2$ operators on another lattice) variables. These collective charge and spin variables commute with each other, and enter into the transformation in a highly non-local way, as shown in Eqs.~\eqref{psidaggerup}--\eqref{psidown}. Being used for correlation functions, the transformation leads to the charge dynamics of the impenetrable anyons, whose statistical angle is averaged out with the weight function defined by spin configurations. 

\textit{Transformation to spin and charge variables.}--- We consider spin-$1/2$ fermions on an infinite one-dimensional lattice. There, $\hat\psi^\dagger_{j\alpha}$, $\hat\psi_{j\alpha}$, and $\hat n_{j\alpha}=\hat\psi^\dagger_{j\alpha}\hat\psi_{j\alpha}$ are the creation, annihilation, and the particle number operators for a site $j$ ($-\infty\le j \le\infty$), and $\alpha=\uparrow,\downarrow$ is the spin index. The local spin vector $\hat{\mathbf{s}}(j) = (\hat s_x(j),\hat s_y(j), \hat s_z(j))$ can be represented as 
\begin{equation} \label{spindef}
\hat{\mathbf{s}}(j) = \frac12 \begin{pmatrix} \hat\psi^\dagger_{j\uparrow} & \hat\psi^\dagger_{j\downarrow} \end{pmatrix} \boldsymbol{\sigma} \begin{pmatrix} \hat\psi_{j\uparrow} \\ \hat\psi_{j\downarrow} \end{pmatrix},
\end{equation}
where $ \boldsymbol{\sigma} = (\sigma_x,\sigma_y, \sigma_z)$ is the vector composed of the three Pauli matrices.  The spin-ladder operators $ \hat s_\pm(j)= \hat s_x(j)\pm i \hat s_y(j)$ read $\hat s_+(j) = \hat\psi^\dagger_{j\uparrow} \hat\psi_{j\downarrow}$ and $\hat s_-(j) = \hat\psi^\dagger_{j\downarrow} \hat\psi_{j\uparrow}$, respectively. We require the total number of fermions in the system, $\hat N= \sum_{j} \hat n_j$, to be a conserved quantity. There could only be either zero or one fermion on each site, 
\begin{equation} \label{njpsi}
\hat n_j \equiv \hat n_{j\uparrow}+ \hat n_{j\downarrow} =\{0,1\},
\end{equation}
due to the NDO constraint.  The projection operator 
\begin{equation} \label{projector}
\hat{\mathcal{X}} = \prod_{j=-\infty}^\infty (1- \hat n_{j\uparrow} \hat n_{j\downarrow})
\end{equation}
applied to the basis state $|\Psi\rangle = \hat\psi^\dagger_{j_1 \alpha_1}\cdots \hat\psi^\dagger_{j_N \alpha_N}|0\rangle$ eliminates those with any number of double occupancies. The remaining ones can be uniquely identified as a product of the states $|f\rangle$ and $|\ell\rangle$:
\begin{equation} \label{psitofl}
|\Psi\rangle = |f\rangle \otimes |\ell\rangle.
\end{equation}
Here, $|f\rangle = \hat c^\dagger_{j_1}\cdots \hat c^\dagger_{j_N}|0\rangle$ is defined by spinless fermions on an infitite lattice placed at the positions of the original spin-$1/2$ fermions. The vacuum $|0\rangle$ for the states $|\Psi\rangle$ and $|f\rangle$ contains no fermions, $\hat\psi_j |0\rangle=0$, and $\hat c_j |0\rangle=0$, respectively. The state $|\ell\rangle = |\alpha_1\cdots\alpha_N\rangle$ of a spin-$1/2$ chain of length $N$ can be represented as $|\ell\rangle=\hat\ell_-(m_1)\cdots \hat\ell_-(m_M)|\Uparrow\rangle$. The set $\{m_1,\ldots,m_M\}$ indicates the positions of the down-spins among $\{\alpha_1,\ldots,\alpha_N\}$, $M$ being the total number of the down-spins. For example, $|\uparrow\downarrow\uparrow\downarrow\downarrow\rangle$ gives $\{m_1,m_2,m_3\}= \{2,4,5\}$. The vacuum $|\Uparrow\rangle$ is the spin-up polarized state. The operator $\hat{\boldsymbol{\ell}}(m)= \boldsymbol{\sigma}(m)/2$ acts on the spin state of the $m$th particle, and  $\hat\ell_\pm=\hat\ell_x\pm i\hat\ell_y$.

We now express spin-$1/2$ fermion fields via operators acting into the spaces formed by $|f\rangle$ and $|\ell\rangle$. The number of particles to the left from the $j$th site is
\begin{equation}
\hat{\mathcal{N}}_j = \sum_{a=-\infty}^j \hat n_a.
\end{equation}
Here, $\hat n_j = \hat c^\dagger_j \hat c_j$ acting onto $|f\rangle$ corresponds to $\hat n_j$ defined by Eq.~\eqref{njpsi}, acting onto $|\Psi\rangle$. Note that the spectrum of the operator $\hat{\mathcal{N}}_j$ is integer-valued. Any operator $\hat{\mathcal{O}}$ depending on $\hat{\mathcal{N}}_j$ can be understood by the following formula:
\begin{equation} \label{ON}
\hat{\mathcal{O}}(\hat{\mathcal{N}}_j) = \sum_{m=-\infty}^\infty \hat{\mathcal{O}}(m) \delta_{m,\hat{\mathcal{N}}_j}.
\end{equation}
The operator $\hat{\mathcal{O}}(m)$ characterizes the state of $m$th particle, and the Kronecker delta
\begin{equation}\label{delta}
\delta_{m,\hat{\mathcal{N}}_j} = \int_0^{2\pi}  \frac{d\lambda}{2\pi}\, e^{ i\lambda(\hat{\mathcal{N}}_j-m)}
\end{equation}
is equal to one for the lattice site at which the $m$th particle is located, and is equal to zero otherwise. The composition law
\begin{equation} \label{composition}
\hat{\mathcal{O}}_1 (\hat{\mathcal{N}}_j) \hat{\mathcal{O}}_2(\hat{\mathcal{N}}_j) = \sum_{m=-\infty}^\infty \hat{\mathcal{O}}_1(m) \hat{\mathcal{O}}_2(m) \delta_{m,\hat{\mathcal{N}}_j}
\end{equation}
stems directly from Eqs.~\eqref{ON} and \eqref{delta}.

We propose the following expressions for the fermion creation operators 
\begin{align}
\hat\psi^\dagger_{j\uparrow} =& \mathcal{P}_{\hat{\mathcal{N}}_j,\hat N}  \hat c^\dagger_j, \label{psidaggerup}\\
\hat \psi^\dagger_{j\downarrow} =& \mathcal{P}_{\hat{\mathcal{N}}_j,\hat N} \hat\ell_-(\hat N) \hat c^\dagger_j . \label{psidaggerdown}
\end{align}
and the corresponding annihilation operators
\begin{align}
\hat\psi_{j\uparrow} =& \hat c_j \hat\eta(\hat N) \mathcal{P}^\dagger_{\hat{\mathcal{N}}_j,\hat N} , \label{psiup}\\
\hat \psi_{j\downarrow} =& \hat c_j \hat\ell_+(\hat N) \mathcal{P}^\dagger_{\hat{\mathcal{N}}_j,\hat N}   . \label{psidown}
\end{align}
The operator $\hat\eta = \hat\ell_+ \hat\ell_- =|\uparrow\rangle\langle\uparrow|$ in Eq.~\eqref{psiup} acts on the site of the spin chain defined by the value of the number operator $\hat N$.
A way to interpret the dependence on $\hat{\mathcal{N}}_j$ is explained by Eqs.~\eqref{ON} and \eqref{delta}. The cyclic shift operator $\mathcal{P}_{m, m^\prime}$ on a lattice encompassing the sites from $m$ to $m^\prime$ is
\begin{equation} \label{permdef}
\mathcal{P}_{m, m^\prime}  = \Pi_{m,m+1} \Pi_{m+1,m+2} \cdots \Pi_{m^\prime -1,m^\prime}.
\end{equation}
The permutation operator $\Pi_{m,m^\prime}$ interchanges the states on the sites $m$ and $m^\prime$, in case of spin-$1/2$ particles it reads
\begin{equation}
\Pi_{m,m^\prime} = \frac12 [\boldsymbol{\sigma}(m) \otimes \boldsymbol{\sigma}(m^\prime) + I\otimes I].
\end{equation}
Here, $I$ is the identity matrix. Evidently, $\Pi$ is its own inverse, $(\Pi_{m,m^\prime})^2=I$, Hermitian, $\Pi^\dagger_{m,m^\prime}=\Pi_{m,m^\prime}$, and unitary, $\Pi^\dagger_{m,m^\prime}\Pi_{m,m^\prime}=I$. This implies $\mathcal{P}_{m^\prime,m} = \mathcal{P}_{m, m^\prime}^{-1} = \mathcal{P}^\dagger_{m, m^\prime}$.
\begin{figure} \label{permfigure}
\includegraphics[width=0.45\textwidth]{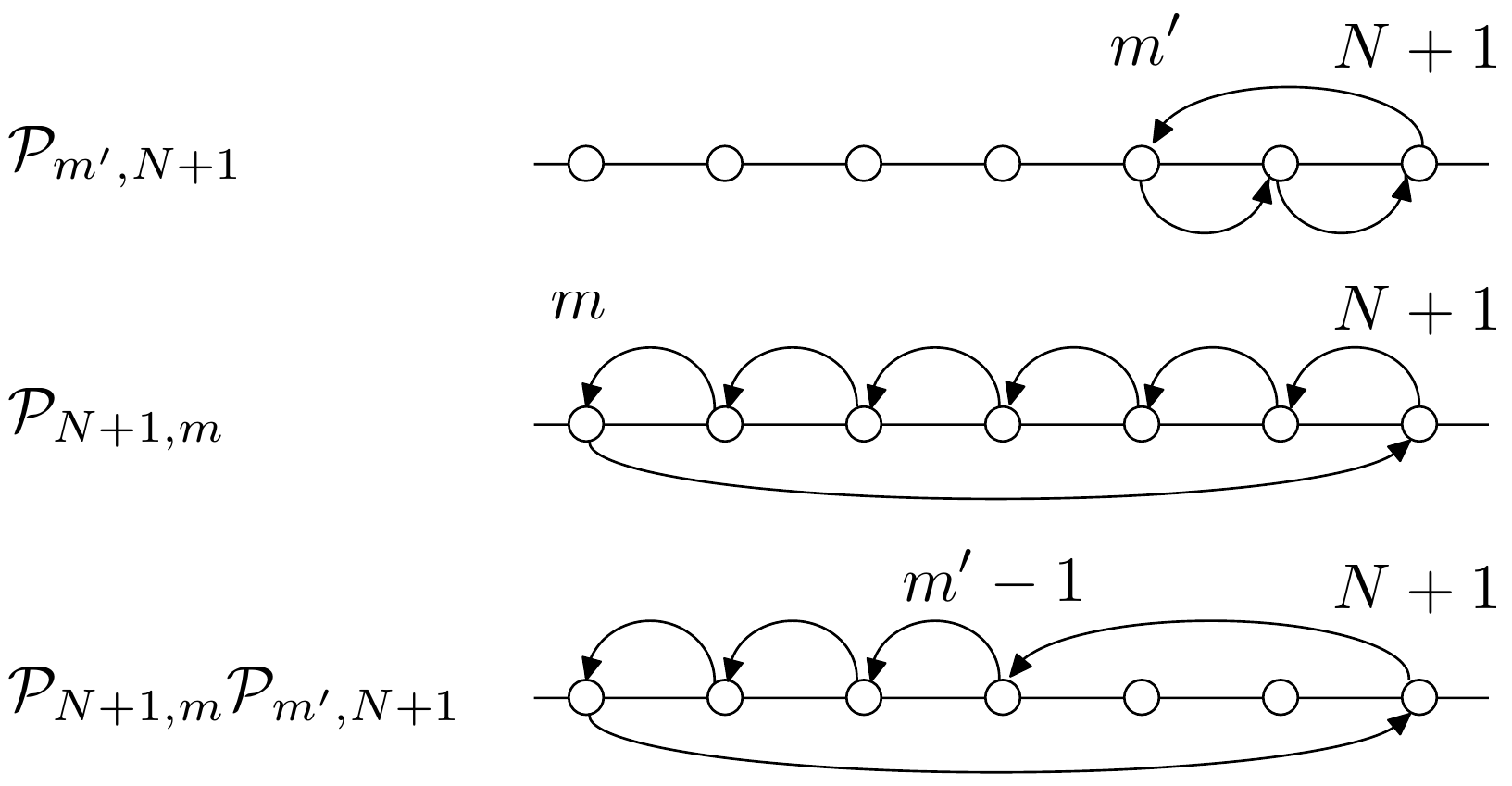}
\caption{\label{fig:permutation} Shown is the action of the operator $\mathcal{P}$ onto the states of the spin chain. The arrows indicate the directions of the transfer of the local states. The outcome of the action of the composition $\mathcal{P}_{N+1,m} \mathcal{P}_{m^\prime,N+1}$ is illustrated for $m^\prime>m$.}
\end{figure}
The action of the operator~\eqref{permdef} onto the states of the spin chain is illustrated in Fig.~\ref{fig:permutation}. Note that the local spin operator~\eqref{spindef} consists of the pairs $\hat\psi^\dagger_{j\alpha} \hat\psi_{j\alpha^\prime}$ where $\hat\psi^\dagger$ and $\hat\psi$ are taken at the same site $j$. As a consequence, the permutation operators cancels out when using Eqs.~\eqref{psidaggerup}--\eqref{psidown}, leading to the representation
\begin{equation} \label{slN}
\hat{\mathbf{s}}(j) = \hat n_j \hat{\boldsymbol{\ell}}(\hat{\mathcal{N}}_j)
\end{equation}
already known in the literature~\cite{zvonarev_BoseHubb_09}. We demonstrate how efficacious are Eqs.~\eqref{psidaggerup}--\eqref{psidown} in revealing the contributions from the spin and charge degrees of freedom into the dynamical correlation functions in the remaining part of the Letter.

\textit{Hamiltonian.}--- We apply the transformations~\eqref{psidaggerup}--\eqref{psidown} to the Hamiltonian
\begin{equation} \label{hu}
\hat H = \hat H_f + \hat H_\ell,
\end{equation}
where
\begin{multline} \label{hcharge}
\hat H_f= \hat{\mathcal{X}} \left[ -t_h \sum_{\substack{j=-\infty \\ \alpha=\uparrow,\downarrow}}^\infty (\hat\psi^\dagger_{j\alpha} \hat\psi_{j+1\alpha} + \textrm{H.c.}) - h \hat{N} \right. \\
\left. +\frac12 \sum_{jj^\prime=-\infty}^\infty :\hat n_j U_{j-j^\prime} \hat n_{j^\prime}: \vphantom{\sum_{\substack{j=-\infty \\ \alpha=\uparrow,\downarrow}}^\infty} \right] \hat{\mathcal{X}}
\end{multline}
is $\mathrm{SU(2)}$-invariant, and the term
\begin{equation} \label{hspin}
\hat H_\ell = 2B \hat{\mathcal{X}} \hat S_z \hat{\mathcal{X}}, \quad \hat S_z = \sum_{j=-\infty}^\infty \hat s_z(j)
\end{equation}
breaks this symmetry due to the magnetic field $B$ applied along the $z$-projection of the total spin. The symbols $\mathrm{H.c.}$ and $:\cdots:$ in Eq.~\eqref{hcharge} stand for the Hermitian conjugate and the normal ordering, respectively. The projection operator $\hat{\mathcal{X}}$, given by Eq.~\eqref{projector}, imposes the NDO constraint. Note that the on-site interaction term $:\hat n_j^2 :U_{0}/2$ implies an infinite energy cost for having two particles on any site in the $U_0\to\infty$ limit. This way, the use of $\hat{\mathcal{X}}$ is equivalent to letting $U_0\to\infty$ in the Hamiltonian~\eqref{hu} with no $\hat{\mathcal{X}}$. The actual value of $U_0$ is irrelevant when $\hat{\mathcal{X}}$ is used, since $\hat{\mathcal{X}} :\hat n_j^2: \hat{\mathcal{X}} =0$.

Using the transformation~\eqref{psidaggerup}--\eqref{psidown} we get Eq.~\eqref{hcharge} written in terms of the spinless fermions exclusively,
\begin{multline} \label{H0_ff}
\hat H_f = -t_h\sum_{j=-\infty}^\infty (\hat c^\dagger_j \hat c_{j+1} +  \textrm{H.c.}) - h \hat N  \\ +\frac12\sum_{j,j^\prime=-\infty}^\infty :\hat n_j U_{j-j^\prime} \hat n_{j^\prime}: 
\end{multline}
and Eq.~\eqref{hspin} containing the spinless fermions as well as the spin operators,
\begin{equation}
\hat H_\ell = 2B \sum_{j=-\infty}^\infty \hat n_j \hat\ell_z(\hat{\mathcal{N}}_j).
\end{equation}
Amazingly, the action of $\hat H_f$ ($\hat H_\ell$) onto the state~\eqref{psitofl} is non-trivial for the $|f\rangle$ ($|\ell\rangle$) part only:
\begin{equation}
\hat H_f |\Psi\rangle = E_f |f\rangle \otimes |\ell\rangle, \quad
\hat H_\ell |\Psi\rangle = |f\rangle \otimes E_\ell |\ell\rangle.
\end{equation}
The energy $E_\ell=2B L_z$, where $L_z$ is the eigenvalue of the operator $\hat L_z= \sum_{m=1}^N \hat\ell_z(m)$, measuring the $z$-projection of the total spin for the state $|\ell\rangle$ of the spin chain. Hence, the spin degeneracy of the Hamiltonian~\eqref{hu} takes place for any $L_z \ne \pm N/2$. Furthermore, $\hat H_\ell=0$ for $B=0$, implying $2^N$-fold degeneracy as long as the system is not put into a finite volume with some boundary conditions.

\textit{Field-field correlation functions in thermal state}.---We consider the one-body correlation functions, describing the particle propagation,
\begin{equation} \label{Gdefp}
G_{p}^\alpha(j-j^\prime,t) = \frac1Z\langle \hat\psi_{j\alpha}(t) \hat\psi^\dagger_{j^\prime\alpha}(0)\rangle_T, \quad \alpha = \uparrow,\downarrow,
\end{equation}
and the hole propagation,
\begin{equation} \label{Gdefh}
G_{h}^\alpha(j-j^\prime,t) = \frac1Z\langle \hat\psi^\dagger_{j\alpha}(t) \hat\psi_{j^\prime\alpha}(0)\rangle_T, \quad \alpha = \uparrow,\downarrow,
\end{equation}
evaluated at temperature $T$, chemical potential $h$, and magnetic field $B$, on a thermals state
\begin{equation}
\langle\cdots\rangle_T = \sum_{N=0}^\infty \sum_{f,\ell} \langle \Psi|e^{-\beta \hat H}\cdots |\Psi\rangle,
\end{equation}
where $|\Psi\rangle$ is given by Eq.~\eqref{psitofl}. The sum over $f$ runs through all possible values of the free-particle momenta characterizing the $N$-fermion state $|f\rangle$. The sum over $\ell$ runs through all possible configurations of the $z$-projection of the spins, $Z$ is the grand partition function, and $\beta=T^{-1}$. The symmetry
\begin{equation}
G^\uparrow_{p(h)}(j-j^\prime,t;h,B) = G^\downarrow_{p(h)}(j-j^\prime,t;h,-B)
\end{equation}
makes it sufficient to evaluate $G^\uparrow$ only.

Using Eqs.~\eqref{ON}--\eqref{psidown} we factorize the matrix element from Eq.~\eqref{Gdefp} into two parts,
\begin{multline}
\langle \Psi| \hat\psi_{j\uparrow}(t) \hat\psi^\dagger_{j^\prime\uparrow}(0) |\Psi\rangle = \sum_{m,m^\prime=-\infty}^\infty  \int_0^{2\pi} \frac{d\lambda}{2\pi} \frac{d\lambda^\prime}{2\pi} \\
e^{-i\lambda m+i\lambda^\prime m^\prime} e^{-\beta (E_f+E_\ell)}\mathcal{C}_p(\lambda,\lambda^\prime;j-j^\prime;t)\mathcal{S}(m,m^\prime).
\end{multline}
The first one encompasses the contributions from the state $|f\rangle$ of spinless fermions,
\begin{equation}\label{cf}
\mathcal{C}_p(\lambda, \lambda^\prime;j-j^\prime;t) = \langle f| \hat c_j(t) e^{i\lambda \hat{\mathcal{N}}_j(t)} e^{-i\lambda^\prime \hat{\mathcal{N}}_{j^\prime}(0)} \hat c^\dagger_{j^\prime}|f\rangle.
\end{equation}
Its non-trivial time evolution is governed by the Hamiltonian~\eqref{H0_ff}. The second one involves the state $|\ell\rangle$ of the spin chain, and the existence of the free fermions is only noticed through their total number $N$, which defines the length of the chain,
\begin{multline}\label{sf}
\mathcal{S}(m,m^\prime) = \langle\ell|\mathcal{P}_{N+1,m} \mathcal{P}_{m^\prime,N+1} |\ell\rangle\\
= \langle\ell| \prod_{j=\min\{m,m^\prime\}}^{\max\{m,m^\prime\}-1} [\frac12 I+\hat\ell_z(j)] |\ell\rangle.
\end{multline}
This part is time-independent, since the cyclic shift operator, Eq.~\eqref{permdef} does not change the value of the $z$-projection of the total spin, $L_z$. The action of the operator $\mathcal{P}_{N+1,m} \mathcal{P}_{m^\prime,N+1}$, illustrated in Fig.~\ref{fig:permutation}, leads to vanishing $\mathcal{S}$ if any spin between the sites $m$ and $m^\prime$ is pointed down. This way we get the right hand side of Eq.~\eqref{sf}.

We proceed further by substituting Eq.~\eqref{sf} into Eq.~\eqref{Gdefp} and taking the sum over the spin configurations,
\begin{equation}
\sum_\ell e^{-\beta E_{\ell}}\mathcal{S}(m,m^\prime) = \frac{\left[2\cosh(\beta B)\right]^N}{\nu^{|m-m^\prime|}} ,
\end{equation}
where $\nu= 1+e^{2\beta B}$. We get
\begin{multline}
G^\uparrow_p(j-j^\prime,t) = \frac1Z \sum_{\{N\}}e^{-\beta \tilde E_f}\int_0^{2\pi} \frac{d\lambda}{2\pi} \frac{d\lambda^\prime}{2\pi}  \\
\times\mathcal{C}_p(\lambda,\lambda^\prime;j-j^\prime;t) \sum_{m,m^\prime=-\infty}^\infty  \frac{e^{-i\lambda m+i\lambda^\prime m^\prime}}{\nu^{|m-m^\prime|}} ,
\end{multline}
where
\begin{equation} \label{Etilde}
\tilde E_f = E_f - \frac1\beta N\ln[2\cosh(\beta B)],
\end{equation}
and the sum over $\{N\}$ encompasses the ones over $N$ and $f$. The partition function $Z$ can be taken over the fermion configurations $f$ with the energies given by Eq.~\eqref{Etilde}. We have
\begin{equation}
\sum_{m,m^\prime=-\infty}^\infty  \frac{e^{-i\lambda m+i\lambda^\prime m^\prime}}{\nu^{|m-m^\prime|}}
=2\pi \delta(\lambda-\lambda^\prime) F(\lambda;T),
\end{equation}
where
\begin{equation} \label{Flambda}
F(\lambda;\nu) = 1+ \sum_{m=1}^\infty \nu^{-m}(e^{im\lambda}+e^{-im\lambda}).
\end{equation}
Therefore,
\begin{equation} \label{Gddfinal}
G^\uparrow_p(j-j^\prime,t) = \int_0^{2\pi} \frac{d\lambda}{2\pi} F(\lambda;\nu) \mathcal{C}_p(\lambda;j-j^\prime;t;T),
\end{equation}
where
\begin{equation} \label{clambdaT}
\mathcal{C}_p(\lambda;j-j^\prime;t;T) = \frac1Z \sum_{\{N\}} e^{-\beta \tilde E_f}\mathcal{C}_p(\lambda;j-j^\prime;t),
\end{equation}
and we write $\mathcal{C}_p(\lambda)$ in place of $\mathcal{C}_p(\lambda,\lambda)$ in order to lighten notations. The summation on the right hand side of Eq.~\eqref{clambdaT} represents the definition of the thermal state for the spinless fermions with the spectum given by $\tilde E_f$.

The hole correlation function~\eqref{Gdefh} is treated the same way as the particle one. The result is given by Eqs.~\eqref{Gddfinal} and~\eqref{clambdaT} with $\mathcal{C}_p$ replaced by
\begin{equation}\label{cfh}
\mathcal{C}_h(\lambda;j-j^\prime;t) = \langle f|  e^{i\lambda \hat{\mathcal{N}}_j(t)} \hat c^\dagger_j(t) \hat c_{j^\prime} e^{-i\lambda\hat{\mathcal{N}}_{j^\prime}(0)} |f\rangle.
\end{equation}

\textit{Emergence of impenetrable anyons}.--- The operator $\hat a_j= \hat c_j e^{-i\lambda \hat{\mathcal{N}}_j}$ satisfies the commutation relations
\begin{align}
& \hat a_j \hat a^\dagger_{j^\prime} + e^{-i\lambda \epsilon(j-j^\prime)} \hat a^\dagger_{j^\prime} \hat a_j   =\delta_{jj^\prime},\\
& \hat a_j \hat a_{j^\prime} + e^{i\lambda \epsilon(j-j^\prime)} \hat a_{j^\prime} \hat a_j =0 ,
\end{align}
where $\epsilon(x)=|x|/x$, and $\epsilon(0)=0$. This is the fermion-anyon mapping discussed in Ref.~\cite{girardeau_anyon_06}. The function $\mathcal{C}_p(\lambda)$ turns into
\begin{equation}
\mathcal{C}_p (-\lambda;j-j^\prime;t)= \langle f| \hat a_j(t) \hat a^\dagger_{j^\prime}(0) |f\rangle,
\end{equation}
which is a correlation function of the impenetrable anyons on a lattice, the variable $\lambda$ being the statistical angle.

The emergence of the anyon correlation function and its subsequent integration over $\lambda$ with the function $F$ in Eq.~\eqref{Gddfinal} could be understood as follows. Let us consider a system with $M$ spin-up and $N-M$ spin-down particles. Pick one spin-up particle among them, and pull it through the whole system, subsequently interchanging its coordinate with those of the other particles. The interchanges with the spin-down particles are non-trivial: the spin part of the wave function could give any phase factor since its symmetry is not restricted by the fermion symmetry of the total wave function. We stress that formalizing our \textit{a posteriori} explanation of the structure of Eq.~\eqref{Gddfinal} by examining exact finite-$N$ wave functions in the coordinate representations (given, for example, in the Refs.~\cite{izergin_impenetrable_fermions_98,izergin_impenetrable_hubbard_98}) goes beyond the scope of the Letter.

\textit{Place among other approaches}.--- The Hamiltonian~\eqref{hu} with $U_{j-j^\prime}=0$ represents the exactly solvable $t-0$ model, also known as the Hubbard model in the limit of infinitely strong repulsion~\cite{essler_book_1DHubbard}. There, Eq.~\eqref{Gddfinal} has been obtained in the form of a Fredholm determinant with the use of the exact wave functions in the coordinate representation~\cite{izergin_impenetrable_bosefermi_short_97,izergin_impenetrable_fermions_98,izergin_impenetrable_hubbard_98}. The transformation~\eqref{psidaggerup}--\eqref{psidown} leading to Eq.~\eqref{Gddfinal}, combined with the ones given in Ref.~\cite{zvonarev_string_09} for the function~\eqref{cf} bring us the same Fredholm determinant representation through much shorter calculations. Note that the model~\eqref{hu} is also exactly solvable when $U_{j-j^\prime}=U\delta_{j,j^\prime\pm 1}$. In this case, the Hamiltonian~\eqref{H0_ff} can be mapped onto the one of the XXZ Heisenberg magnet, and the function~\eqref{cf} can, in principle, be calculated by the Bethe Ansatz method.

Special attention has been paid in the literature to the model in the $T\to 0$ limit. Its ground state is non-degenerate and spin-up (-down) polarized for $B$ negative (positive). In the former case, Eq.~\eqref{Gddfinal} describes a spin-up fermion propagating through a gas of the other spin-up fermions \footnote{For dynamic and finite temperature aspects of such a ``mobile impurity'' see also \cite{gamayun_impurity_Green_FTG_14,gamayun_impurity_Green_FTG_16,gamayun_protocol_impurity_18,10.21468/SciPostPhys.8.4.053,Gamayun2022}}. We have $F=2\pi \delta(\lambda)$ in Eq.~\eqref{Flambda}, hence $G^\uparrow_p = \langle \hat c_j(t) \hat c^\dagger_{j^\prime}\rangle$. In the latter case, Eq.~\eqref{Gddfinal} describes a spin-up fermion (an impurity particle) propagating through a gas of spin-down fermions. We have $F=1$, and the long time and distance asymptotic behaviour of $G^\uparrow_p$ reveals the logarithmic diffusion phenomenon~\cite{zvonarev_ferrobosons_07, akhanjee_ferrobosons_07}. The non-degeneracy of the ground state at $B\ne 0$ stands in a sharp contrast to the high degeneracy at $B=0$, where  $F$ is given by Eq.~\eqref{Flambda} with $\nu=2$. This regime is known as the spin-incoherent one~\cite{cheianov_spin_decoherent_short_04, fiete_spin_decoherent_04, fiete_SI_07}. A challenge put forward in the aforementioned works was to find a low-energy effective field theory, since the low-enegry spectrum of spin excitations cannot be linearized for $B>0$ and $B=0$, and the LL theory is inapplicable. The representation~\eqref{Gddfinal} resolves this problem in the following way: the LL theory in applicable to the function $\mathcal{C}_p$; the spin excitations are accounted for by the integral over $\lambda$ with the weight function $F$ without any approximation, which is equivalent to counting the number of worldlines within the first-quantized path integral approach implemented in Refs.~\cite{zvonarev_ferrobosons_07, fiete_spin_decoherent_04}.

\section{Acknowledgements}
\noindent We thank V. Cheianov and K. Seetharam for fruitful discussions. O.G. acknowledges support from the Polish National Agency for Academic Exchange (NAWA) through the Grant No. PPN/ULM/2020/1/00247. O.G. is grateful to Galileo Galilei Institute for hospitality and support during the scientific program on “Randomness, Integrability, and Universality”, where part of this work was done. The work of E.~Q. is supported by Grant No.~ANR-16-CE91-0009-01. K.B. thanks S.~Bocini, V.~Marić, L.~Zadnik and M.~Fagotti for useful discussions. The work of K.B. was partially supported by the European Research Council under the Starting Grant No. 805252 LoCoMacro. The work of M.~B.~Z. is supported by Grant No.~ANR-16-CE91-0009-01 and CNRS grant PICS06738. M.~B.~Z. acknowledges Russian Quantum Center and Prof. A.~Fedorov for their hospitality during the work.

\bibliography{zvonarev}

\begin{thebibliography}{32}%
\makeatletter
\providecommand \@ifxundefined [1]{%
 \@ifx{#1\undefined}
}%
\providecommand \@ifnum [1]{%
 \ifnum #1\expandafter \@firstoftwo
 \else \expandafter \@secondoftwo
 \fi
}%
\providecommand \@ifx [1]{%
 \ifx #1\expandafter \@firstoftwo
 \else \expandafter \@secondoftwo
 \fi
}%
\providecommand \natexlab [1]{#1}%
\providecommand \enquote  [1]{``#1''}%
\providecommand \bibnamefont  [1]{#1}%
\providecommand \bibfnamefont [1]{#1}%
\providecommand \citenamefont [1]{#1}%
\providecommand \href@noop [0]{\@secondoftwo}%
\providecommand \href [0]{\begingroup \@sanitize@url \@href}%
\providecommand \@href[1]{\@@startlink{#1}\@@href}%
\providecommand \@@href[1]{\endgroup#1\@@endlink}%
\providecommand \@sanitize@url [0]{\catcode `\\12\catcode `\$12\catcode
  `\&12\catcode `\#12\catcode `\^12\catcode `\_12\catcode `\%12\relax}%
\providecommand \@@startlink[1]{}%
\providecommand \@@endlink[0]{}%
\providecommand \url  [0]{\begingroup\@sanitize@url \@url }%
\providecommand \@url [1]{\endgroup\@href {#1}{\urlprefix }}%
\providecommand \urlprefix  [0]{URL }%
\providecommand \Eprint [0]{\href }%
\providecommand \doibase [0]{https://doi.org/}%
\providecommand \selectlanguage [0]{\@gobble}%
\providecommand \bibinfo  [0]{\@secondoftwo}%
\providecommand \bibfield  [0]{\@secondoftwo}%
\providecommand \translation [1]{[#1]}%
\providecommand \BibitemOpen [0]{}%
\providecommand \bibitemStop [0]{}%
\providecommand \bibitemNoStop [0]{.\EOS\space}%
\providecommand \EOS [0]{\spacefactor3000\relax}%
\providecommand \BibitemShut  [1]{\csname bibitem#1\endcsname}%
\let\auto@bib@innerbib\@empty
\bibitem [{\citenamefont {Pitaevskii}\ and\ \citenamefont
  {Stringari}(2003)}]{pitaevskii_book_BEC}%
  \BibitemOpen
  \bibfield  {author} {\bibinfo {author} {\bibfnamefont {L.}~\bibnamefont
  {Pitaevskii}}\ and\ \bibinfo {author} {\bibfnamefont {S.}~\bibnamefont
  {Stringari}},\ }\href@noop {} {\emph {\bibinfo {title} {Bose-Einstein
  Condensation}}}\ (\bibinfo  {publisher} {Oxford University Press},\ \bibinfo
  {address} {Oxford},\ \bibinfo {year} {2003})\BibitemShut {NoStop}%
\bibitem [{\citenamefont {Girardeau}(1960)}]{girardeau_impurity_TG_60}%
  \BibitemOpen
  \bibfield  {author} {\bibinfo {author} {\bibfnamefont {M.}~\bibnamefont
  {Girardeau}},\ }\bibfield  {title} {\bibinfo {title} {Relationship between
  systems of impenetrable bosons and fermions in one dimension},\ }\href
  {https://doi.org/10.1063/1.1703687} {\bibfield  {journal} {\bibinfo
  {journal} {J. Math. Phys.}\ }\textbf {\bibinfo {volume} {1}},\ \bibinfo
  {pages} {516} (\bibinfo {year} {1960})}\BibitemShut {NoStop}%
\bibitem [{\citenamefont {Gogolin}\ \emph {et~al.}(1999)\citenamefont
  {Gogolin}, \citenamefont {Nersesyan},\ and\ \citenamefont
  {Tsvelik}}]{gogolin_1dbook}%
  \BibitemOpen
  \bibfield  {author} {\bibinfo {author} {\bibfnamefont {A.~O.}\ \bibnamefont
  {Gogolin}}, \bibinfo {author} {\bibfnamefont {A.~A.}\ \bibnamefont
  {Nersesyan}},\ and\ \bibinfo {author} {\bibfnamefont {A.~M.}\ \bibnamefont
  {Tsvelik}},\ }\href@noop {} {\emph {\bibinfo {title} {Bosonization and
  Strongly Correlated Systems}}}\ (\bibinfo  {publisher} {Cambridge University
  Press},\ \bibinfo {address} {Cambridge},\ \bibinfo {year} {1999})\BibitemShut
  {NoStop}%
\bibitem [{\citenamefont {Giamarchi}(2004)}]{giamarchi_book_1d}%
  \BibitemOpen
  \bibfield  {author} {\bibinfo {author} {\bibfnamefont {T.}~\bibnamefont
  {Giamarchi}},\ }\href@noop {} {\emph {\bibinfo {title} {Quantum Physics in
  One Dimension}}}\ (\bibinfo  {publisher} {Oxford University Press},\ \bibinfo
  {address} {Oxford},\ \bibinfo {year} {2004})\BibitemShut {NoStop}%
\bibitem [{\citenamefont {Cheianov}\ and\ \citenamefont
  {Zvonarev}(2004)}]{cheianov_spin_decoherent_short_04}%
  \BibitemOpen
  \bibfield  {author} {\bibinfo {author} {\bibfnamefont {V.~V.}\ \bibnamefont
  {Cheianov}}\ and\ \bibinfo {author} {\bibfnamefont {M.~B.}\ \bibnamefont
  {Zvonarev}},\ }\bibfield  {title} {\bibinfo {title} {{Nonunitary Spin-Charge
  Separation in a One-Dimensional Fermion Gas}},\ }\href
  {https://doi.org/10.1103/PhysRevLett.92.176401} {\bibfield  {journal}
  {\bibinfo  {journal} {Phys. Rev. Lett.}\ }\textbf {\bibinfo {volume} {92}},\
  \bibinfo {pages} {176401} (\bibinfo {year} {2004})},\ \Eprint
  {https://arxiv.org/abs/arXiv:cond-mat/0308470} {arXiv:cond-mat/0308470}
  \BibitemShut {NoStop}%
\bibitem [{\citenamefont {Fiete}\ and\ \citenamefont
  {Balents}(2004)}]{fiete_spin_decoherent_04}%
  \BibitemOpen
  \bibfield  {author} {\bibinfo {author} {\bibfnamefont {G.~A.}\ \bibnamefont
  {Fiete}}\ and\ \bibinfo {author} {\bibfnamefont {L.}~\bibnamefont
  {Balents}},\ }\bibfield  {title} {\bibinfo {title} {{Green's Function for
  Magnetically Incoherent Interacting Electrons in One Dimension}},\ }\href
  {https://doi.org/10.1103/PhysRevLett.93.226401} {\bibfield  {journal}
  {\bibinfo  {journal} {Phys. Rev. Lett.}\ }\textbf {\bibinfo {volume} {93}},\
  \bibinfo {pages} {226401} (\bibinfo {year} {2004})},\ \Eprint
  {https://arxiv.org/abs/arXiv:cond-mat/0403744} {arXiv:cond-mat/0403744}
  \BibitemShut {NoStop}%
\bibitem [{\citenamefont {Fiete}(2007)}]{fiete_SI_07}%
  \BibitemOpen
  \bibfield  {author} {\bibinfo {author} {\bibfnamefont {G.~A.}\ \bibnamefont
  {Fiete}},\ }\bibfield  {title} {\bibinfo {title} {{Colloquium: The
  spin-incoherent Luttinger liquid}},\ }\href
  {https://doi.org/10.1103/RevModPhys.79.801} {\bibfield  {journal} {\bibinfo
  {journal} {Rev. Mod. Phys.}\ }\textbf {\bibinfo {volume} {79}},\ \bibinfo
  {pages} {801} (\bibinfo {year} {2007})},\ \Eprint
  {https://arxiv.org/abs/arXiv:cond-mat/0611597} {arXiv:cond-mat/0611597}
  \BibitemShut {NoStop}%
\bibitem [{\citenamefont {Zvonarev}\ \emph {et~al.}(2007)\citenamefont
  {Zvonarev}, \citenamefont {Cheianov},\ and\ \citenamefont
  {Giamarchi}}]{zvonarev_ferrobosons_07}%
  \BibitemOpen
  \bibfield  {author} {\bibinfo {author} {\bibfnamefont {M.~B.}\ \bibnamefont
  {Zvonarev}}, \bibinfo {author} {\bibfnamefont {V.~V.}\ \bibnamefont
  {Cheianov}},\ and\ \bibinfo {author} {\bibfnamefont {T.}~\bibnamefont
  {Giamarchi}},\ }\bibfield  {title} {\bibinfo {title} {{Spin dynamics in a
  one-dimensional ferromagnetic Bose gas}},\ }\href
  {https://doi.org/10.1103/PhysRevLett.99.240404} {\bibfield  {journal}
  {\bibinfo  {journal} {Phys. Rev. Lett.}\ }\textbf {\bibinfo {volume} {99}},\
  \bibinfo {pages} {240404} (\bibinfo {year} {2007})},\ \Eprint
  {https://arxiv.org/abs/arXiv:0708.3638} {arXiv:0708.3638} \BibitemShut
  {NoStop}%
\bibitem [{\citenamefont {Akhanjee}\ and\ \citenamefont
  {Tserkovnyak}(2007)}]{akhanjee_ferrobosons_07}%
  \BibitemOpen
  \bibfield  {author} {\bibinfo {author} {\bibfnamefont {S.}~\bibnamefont
  {Akhanjee}}\ and\ \bibinfo {author} {\bibfnamefont {Y.}~\bibnamefont
  {Tserkovnyak}},\ }\bibfield  {title} {\bibinfo {title} {Spin-charge
  separation in a strongly correlated spin-polarized chain},\ }\href
  {https://doi.org/10.1103/PhysRevB.76.140408} {\bibfield  {journal} {\bibinfo
  {journal} {Phys. Rev. B}\ }\textbf {\bibinfo {volume} {76}},\ \bibinfo
  {pages} {140408(R)} (\bibinfo {year} {2007})},\ \Eprint
  {https://arxiv.org/abs/arXiv:0708.4012} {arXiv:0708.4012} \BibitemShut
  {NoStop}%
\bibitem [{\citenamefont {Kamenev}\ and\ \citenamefont
  {Glazman}(2009)}]{kamenev_spinor_bosons_09}%
  \BibitemOpen
  \bibfield  {author} {\bibinfo {author} {\bibfnamefont {A.}~\bibnamefont
  {Kamenev}}\ and\ \bibinfo {author} {\bibfnamefont {L.~I.}\ \bibnamefont
  {Glazman}},\ }\bibfield  {title} {\bibinfo {title} {{Dynamics of a
  one-dimensional spinor Bose liquid: A phenomenological approach}},\ }\href
  {https://doi.org/10.1103/PhysRevA.80.011603} {\bibfield  {journal} {\bibinfo
  {journal} {Phys. Rev. A}\ }\textbf {\bibinfo {volume} {80}},\ \bibinfo
  {pages} {011603} (\bibinfo {year} {2009})},\ \Eprint
  {https://arxiv.org/abs/arXiv:0808.0479} {arXiv:0808.0479} \BibitemShut
  {NoStop}%
\bibitem [{\citenamefont {Zvonarev}\ \emph
  {et~al.}(2009{\natexlab{a}})\citenamefont {Zvonarev}, \citenamefont
  {Cheianov},\ and\ \citenamefont {Giamarchi}}]{zvonarev_Yang_Gaudin_09}%
  \BibitemOpen
  \bibfield  {author} {\bibinfo {author} {\bibfnamefont {M.~B.}\ \bibnamefont
  {Zvonarev}}, \bibinfo {author} {\bibfnamefont {V.~V.}\ \bibnamefont
  {Cheianov}},\ and\ \bibinfo {author} {\bibfnamefont {T.}~\bibnamefont
  {Giamarchi}},\ }\bibfield  {title} {\bibinfo {title} {{Edge exponent in the
  dynamic spin structure factor of the Yang-Gaudin model}},\ }\href
  {https://doi.org/10.1103/PhysRevB.80.201102} {\bibfield  {journal} {\bibinfo
  {journal} {Phys. Rev. B}\ }\textbf {\bibinfo {volume} {80}},\ \bibinfo
  {pages} {201102} (\bibinfo {year} {2009}{\natexlab{a}})},\ \Eprint
  {https://arxiv.org/abs/0905.0598} {0905.0598} \BibitemShut {NoStop}%
\bibitem [{\citenamefont {Zvonarev}\ \emph
  {et~al.}(2009{\natexlab{b}})\citenamefont {Zvonarev}, \citenamefont
  {Cheianov},\ and\ \citenamefont {Giamarchi}}]{zvonarev_BoseHubb_09}%
  \BibitemOpen
  \bibfield  {author} {\bibinfo {author} {\bibfnamefont {M.~B.}\ \bibnamefont
  {Zvonarev}}, \bibinfo {author} {\bibfnamefont {V.~V.}\ \bibnamefont
  {Cheianov}},\ and\ \bibinfo {author} {\bibfnamefont {T.}~\bibnamefont
  {Giamarchi}},\ }\bibfield  {title} {\bibinfo {title} {{Dynamical Properties
  of the One-Dimensional Spin-1/2 Bose-Hubbard Model near a Mott-Insulator to
  Ferromagnetic-Liquid Transition}},\ }\href
  {https://doi.org/10.1103/PhysRevLett.103.110401} {\bibfield  {journal}
  {\bibinfo  {journal} {Phys. Rev. Lett.}\ }\textbf {\bibinfo {volume} {103}},\
  \bibinfo {pages} {110401} (\bibinfo {year} {2009}{\natexlab{b}})},\ \Eprint
  {https://arxiv.org/abs/arXiv:0811.2676} {arXiv:0811.2676} \BibitemShut
  {NoStop}%
\bibitem [{\citenamefont {Imambekov}\ \emph {et~al.}(2012)\citenamefont
  {Imambekov}, \citenamefont {Schmidt},\ and\ \citenamefont
  {Glazman}}]{imambekov_review_12}%
  \BibitemOpen
  \bibfield  {author} {\bibinfo {author} {\bibfnamefont {A.}~\bibnamefont
  {Imambekov}}, \bibinfo {author} {\bibfnamefont {T.~L.}\ \bibnamefont
  {Schmidt}},\ and\ \bibinfo {author} {\bibfnamefont {L.~I.}\ \bibnamefont
  {Glazman}},\ }\bibfield  {title} {\bibinfo {title} {{One-dimensional quantum
  liquids: Beyond the Luttinger liquid paradigm}},\ }\href
  {https://doi.org/10.1103/RevModPhys.84.1253} {\bibfield  {journal} {\bibinfo
  {journal} {Rev. Mod. Phys.}\ }\textbf {\bibinfo {volume} {84}},\ \bibinfo
  {pages} {1253} (\bibinfo {year} {2012})},\ \Eprint
  {https://arxiv.org/abs/arXiv:1110.1374} {arXiv:1110.1374} \BibitemShut
  {NoStop}%
\bibitem [{\citenamefont {Senaratne}\ \emph {et~al.}(2022)\citenamefont
  {Senaratne}, \citenamefont {Cavazos-Cavazos}, \citenamefont {Wang},
  \citenamefont {He}, \citenamefont {Chang}, \citenamefont {Kafle},
  \citenamefont {Pu}, \citenamefont {Guan},\ and\ \citenamefont
  {Hulet}}]{senarathe_spincharge_fermi_22}%
  \BibitemOpen
  \bibfield  {author} {\bibinfo {author} {\bibfnamefont {R.}~\bibnamefont
  {Senaratne}}, \bibinfo {author} {\bibfnamefont {D.}~\bibnamefont
  {Cavazos-Cavazos}}, \bibinfo {author} {\bibfnamefont {S.}~\bibnamefont
  {Wang}}, \bibinfo {author} {\bibfnamefont {F.}~\bibnamefont {He}}, \bibinfo
  {author} {\bibfnamefont {Y.-T.}\ \bibnamefont {Chang}}, \bibinfo {author}
  {\bibfnamefont {A.}~\bibnamefont {Kafle}}, \bibinfo {author} {\bibfnamefont
  {H.}~\bibnamefont {Pu}}, \bibinfo {author} {\bibfnamefont {X.-W.}\
  \bibnamefont {Guan}},\ and\ \bibinfo {author} {\bibfnamefont {R.~G.}\
  \bibnamefont {Hulet}},\ }\bibfield  {title} {\bibinfo {title} {{Spin-charge
  separation in a one-dimensional Fermi gas with tunable interactions}},\
  }\href {https://doi.org/10.1126/science.abn1719} {\bibfield  {journal}
  {\bibinfo  {journal} {Science}\ }\textbf {\bibinfo {volume} {376}},\ \bibinfo
  {pages} {1305} (\bibinfo {year} {2022})},\ \Eprint
  {https://arxiv.org/abs/arXiv:2111.11545} {arXiv:2111.11545} \BibitemShut
  {NoStop}%
\bibitem [{\citenamefont {Hilker}\ \emph {et~al.}(2017)\citenamefont {Hilker},
  \citenamefont {Salomon}, \citenamefont {Grusdt}, \citenamefont {Omran},
  \citenamefont {Boll}, \citenamefont {Demler}, \citenamefont {Bloch},\ and\
  \citenamefont {Gross}}]{hilker_string_17}%
  \BibitemOpen
  \bibfield  {author} {\bibinfo {author} {\bibfnamefont {T.~A.}\ \bibnamefont
  {Hilker}}, \bibinfo {author} {\bibfnamefont {G.}~\bibnamefont {Salomon}},
  \bibinfo {author} {\bibfnamefont {F.}~\bibnamefont {Grusdt}}, \bibinfo
  {author} {\bibfnamefont {A.}~\bibnamefont {Omran}}, \bibinfo {author}
  {\bibfnamefont {M.}~\bibnamefont {Boll}}, \bibinfo {author} {\bibfnamefont
  {E.}~\bibnamefont {Demler}}, \bibinfo {author} {\bibfnamefont
  {I.}~\bibnamefont {Bloch}},\ and\ \bibinfo {author} {\bibfnamefont
  {C.}~\bibnamefont {Gross}},\ }\bibfield  {title} {\bibinfo {title}
  {{Revealing hidden antiferromagnetic correlations in doped Hubbard chains via
  string correlators}},\ }\href {https://doi.org/10.1126/science.aam8990}
  {\bibfield  {journal} {\bibinfo  {journal} {Science}\ }\textbf {\bibinfo
  {volume} {357}},\ \bibinfo {pages} {484} (\bibinfo {year} {2017})},\ \Eprint
  {https://arxiv.org/abs/arXiv:1702.00642} {arXiv:1702.00642} \BibitemShut
  {NoStop}%
\bibitem [{\citenamefont {Salomon}\ \emph {et~al.}(2019)\citenamefont
  {Salomon}, \citenamefont {Koepsell}, \citenamefont {Vijayan}, \citenamefont
  {Hilker}, \citenamefont {Nespolo}, \citenamefont {Pollet}, \citenamefont
  {Bloch},\ and\ \citenamefont {Gross}}]{salomon_incommensurate_19}%
  \BibitemOpen
  \bibfield  {author} {\bibinfo {author} {\bibfnamefont {G.}~\bibnamefont
  {Salomon}}, \bibinfo {author} {\bibfnamefont {J.}~\bibnamefont {Koepsell}},
  \bibinfo {author} {\bibfnamefont {J.}~\bibnamefont {Vijayan}}, \bibinfo
  {author} {\bibfnamefont {T.~A.}\ \bibnamefont {Hilker}}, \bibinfo {author}
  {\bibfnamefont {J.}~\bibnamefont {Nespolo}}, \bibinfo {author} {\bibfnamefont
  {L.}~\bibnamefont {Pollet}}, \bibinfo {author} {\bibfnamefont
  {I.}~\bibnamefont {Bloch}},\ and\ \bibinfo {author} {\bibfnamefont
  {C.}~\bibnamefont {Gross}},\ }\bibfield  {title} {\bibinfo {title} {{Direct
  observation of incommensurate magnetism in Hubbard chains}},\ }\href
  {https://doi.org/https://doi.org/10.1038/s41586-018-0778-7} {\bibfield
  {journal} {\bibinfo  {journal} {Nature}\ }\textbf {\bibinfo {volume} {565}},\
  \bibinfo {pages} {56} (\bibinfo {year} {2019})},\ \Eprint
  {https://arxiv.org/abs/arXiv:1803.08892} {arXiv:1803.08892} \BibitemShut
  {NoStop}%
\bibitem [{\citenamefont {Vijayan}\ \emph {et~al.}(2020)\citenamefont
  {Vijayan}, \citenamefont {Sompet}, \citenamefont {Salomon}, \citenamefont
  {Koepsell}, \citenamefont {Hirthe}, \citenamefont {Bohrdt}, \citenamefont
  {Grusdt}, \citenamefont {Bloch},\ and\ \citenamefont
  {Gross}}]{vijayan_deconf_Hubbard_20}%
  \BibitemOpen
  \bibfield  {author} {\bibinfo {author} {\bibfnamefont {J.}~\bibnamefont
  {Vijayan}}, \bibinfo {author} {\bibfnamefont {P.}~\bibnamefont {Sompet}},
  \bibinfo {author} {\bibfnamefont {G.}~\bibnamefont {Salomon}}, \bibinfo
  {author} {\bibfnamefont {J.}~\bibnamefont {Koepsell}}, \bibinfo {author}
  {\bibfnamefont {S.}~\bibnamefont {Hirthe}}, \bibinfo {author} {\bibfnamefont
  {A.}~\bibnamefont {Bohrdt}}, \bibinfo {author} {\bibfnamefont
  {F.}~\bibnamefont {Grusdt}}, \bibinfo {author} {\bibfnamefont
  {I.}~\bibnamefont {Bloch}},\ and\ \bibinfo {author} {\bibfnamefont
  {C.}~\bibnamefont {Gross}},\ }\bibfield  {title} {\bibinfo {title}
  {{Time-resolved observation of spin-charge deconfinement in fermionic Hubbard
  chains}},\ }\href {https://doi.org/10.1126/science.aay2354} {\bibfield
  {journal} {\bibinfo  {journal} {Science}\ }\textbf {\bibinfo {volume}
  {367}},\ \bibinfo {pages} {186} (\bibinfo {year} {2020})},\ \Eprint
  {https://arxiv.org/abs/arXiv:1905.13638} {arXiv:1905.13638} \BibitemShut
  {NoStop}%
\bibitem [{\citenamefont {Ogata}\ and\ \citenamefont
  {Shiba}(1990)}]{ogata_BA_90}%
  \BibitemOpen
  \bibfield  {author} {\bibinfo {author} {\bibfnamefont {M.}~\bibnamefont
  {Ogata}}\ and\ \bibinfo {author} {\bibfnamefont {H.}~\bibnamefont {Shiba}},\
  }\bibfield  {title} {\bibinfo {title} {{Bethe-ansatz wave function, momentum
  distribution, and spin correlation in the one-dimensional strongly correlated
  Hubbard model}},\ }\href {https://doi.org/10.1103/PhysRevB.41.2326}
  {\bibfield  {journal} {\bibinfo  {journal} {Phys. Rev. B}\ }\textbf {\bibinfo
  {volume} {41}},\ \bibinfo {pages} {2326} (\bibinfo {year}
  {1990})}\BibitemShut {NoStop}%
\bibitem [{\citenamefont {Zabrodin}\ and\ \citenamefont
  {Ovchinnikov}(1989)}]{zabrodin_Fermi_strong_89}%
  \BibitemOpen
  \bibfield  {author} {\bibinfo {author} {\bibfnamefont {A.~V.}\ \bibnamefont
  {Zabrodin}}\ and\ \bibinfo {author} {\bibfnamefont {A.~A.}\ \bibnamefont
  {Ovchinnikov}},\ }\bibfield  {title} {\bibinfo {title} {{Spin-density
  correlator of a one-dimensional Fermi gas with strong interaction}},\
  }\href@noop {} {\bibfield  {journal} {\bibinfo  {journal} {Soviet Physics -
  JETP (English Translation)}\ }\textbf {\bibinfo {volume} {69}},\ \bibinfo
  {pages} {750} (\bibinfo {year} {1989})}\BibitemShut {NoStop}%
\bibitem [{\citenamefont {Zabrodin}\ and\ \citenamefont
  {Ovchinnikov}(1990)}]{zabrodin_single_particle_90}%
  \BibitemOpen
  \bibfield  {author} {\bibinfo {author} {\bibfnamefont {A.~V.}\ \bibnamefont
  {Zabrodin}}\ and\ \bibinfo {author} {\bibfnamefont {A.~A.}\ \bibnamefont
  {Ovchinnikov}},\ }\bibfield  {title} {\bibinfo {title} {{Single-particle
  density matrix of a one-dimensional system of spin 1/2 Fermi particles}},\
  }\href {https://doi.org/10.1007/BF01018409} {\bibfield  {journal} {\bibinfo
  {journal} {Theor. Math. Phys.}\ }\textbf {\bibinfo {volume} {85}},\ \bibinfo
  {pages} {1321} (\bibinfo {year} {1990})}\BibitemShut {NoStop}%
\bibitem [{\citenamefont {Izergin}\ \emph {et~al.}(1998)\citenamefont
  {Izergin}, \citenamefont {Pronko},\ and\ \citenamefont
  {Abarenkova}}]{izergin_impenetrable_hubbard_98}%
  \BibitemOpen
  \bibfield  {author} {\bibinfo {author} {\bibfnamefont {A.~G.}\ \bibnamefont
  {Izergin}}, \bibinfo {author} {\bibfnamefont {A.~G.}\ \bibnamefont
  {Pronko}},\ and\ \bibinfo {author} {\bibfnamefont {N.~I.}\ \bibnamefont
  {Abarenkova}},\ }\bibfield  {title} {\bibinfo {title} {{Temperature
  correlators in the one-dimensional Hubbard model in the strong coupling
  limit}},\ }\href {https://doi.org/10.1016/S0375-9601(98)00442-3} {\bibfield
  {journal} {\bibinfo  {journal} {Phys. Lett. A}\ }\textbf {\bibinfo {volume}
  {245}},\ \bibinfo {pages} {537} (\bibinfo {year} {1998})},\ \Eprint
  {https://arxiv.org/abs/hep-th/9801167} {hep-th/9801167} \BibitemShut
  {NoStop}%
\bibitem [{\citenamefont {Girardeau}(2006)}]{girardeau_anyon_06}%
  \BibitemOpen
  \bibfield  {author} {\bibinfo {author} {\bibfnamefont {M.~D.}\ \bibnamefont
  {Girardeau}},\ }\bibfield  {title} {\bibinfo {title} {Anyon-fermion mapping
  and applications to ultracold gases in tight waveguides},\ }\href
  {https://doi.org/10.1103/PhysRevLett.97.100402} {\bibfield  {journal}
  {\bibinfo  {journal} {Phys. Rev. Lett.}\ }\textbf {\bibinfo {volume} {97}},\
  \bibinfo {pages} {100402} (\bibinfo {year} {2006})},\ \Eprint
  {https://arxiv.org/abs/arXiv:cond-mat/0604357} {arXiv:cond-mat/0604357}
  \BibitemShut {NoStop}%
\bibitem [{\citenamefont {Izergin}\ and\ \citenamefont
  {Pronko}(1998)}]{izergin_impenetrable_fermions_98}%
  \BibitemOpen
  \bibfield  {author} {\bibinfo {author} {\bibfnamefont {A.~G.}\ \bibnamefont
  {Izergin}}\ and\ \bibinfo {author} {\bibfnamefont {A.~G.}\ \bibnamefont
  {Pronko}},\ }\bibfield  {title} {\bibinfo {title} {Temperature correlators in
  the two-component one-dimensional gas},\ }\href
  {https://doi.org/10.1016/S0550-3213(98)00182-5} {\bibfield  {journal}
  {\bibinfo  {journal} {Nucl. Phys. B}\ }\textbf {\bibinfo {volume} {520}},\
  \bibinfo {pages} {594} (\bibinfo {year} {1998})},\ \Eprint
  {https://arxiv.org/abs/arXiv:solv-int/9801004} {arXiv:solv-int/9801004}
  \BibitemShut {NoStop}%
\bibitem [{\citenamefont {Essler}\ \emph {et~al.}(2005)\citenamefont {Essler},
  \citenamefont {Frahm}, \citenamefont {G\"ohmann}, \citenamefont {Kl\"umper},\
  and\ \citenamefont {Korepin}}]{essler_book_1DHubbard}%
  \BibitemOpen
  \bibfield  {author} {\bibinfo {author} {\bibfnamefont {F.~H.~L.}\
  \bibnamefont {Essler}}, \bibinfo {author} {\bibfnamefont {H.}~\bibnamefont
  {Frahm}}, \bibinfo {author} {\bibfnamefont {F.}~\bibnamefont {G\"ohmann}},
  \bibinfo {author} {\bibfnamefont {A.}~\bibnamefont {Kl\"umper}},\ and\
  \bibinfo {author} {\bibfnamefont {V.~E.}\ \bibnamefont {Korepin}},\
  }\href@noop {} {\emph {\bibinfo {title} {The One-Dimensional Hubbard
  Model}}}\ (\bibinfo  {publisher} {Cambridge University Press},\ \bibinfo
  {address} {Cambridge},\ \bibinfo {year} {2005})\BibitemShut {NoStop}%
\bibitem [{\citenamefont {Izergin}\ and\ \citenamefont
  {Pronko}(1997)}]{izergin_impenetrable_bosefermi_short_97}%
  \BibitemOpen
  \bibfield  {author} {\bibinfo {author} {\bibfnamefont {A.~G.}\ \bibnamefont
  {Izergin}}\ and\ \bibinfo {author} {\bibfnamefont {A.~G.}\ \bibnamefont
  {Pronko}},\ }\bibfield  {title} {\bibinfo {title} {{Correlators in the
  one-dimensional two-component Bose and Fermi gases}},\ }\href
  {https://doi.org/10.1016/S0375-9601(97)00791-3} {\bibfield  {journal}
  {\bibinfo  {journal} {Phys. Lett. A}\ }\textbf {\bibinfo {volume} {236}},\
  \bibinfo {pages} {445} (\bibinfo {year} {1997})}\BibitemShut {NoStop}%
\bibitem [{\citenamefont {Zvonarev}\ \emph
  {et~al.}(2009{\natexlab{c}})\citenamefont {Zvonarev}, \citenamefont
  {Cheianov},\ and\ \citenamefont {Giamarchi}}]{zvonarev_string_09}%
  \BibitemOpen
  \bibfield  {author} {\bibinfo {author} {\bibfnamefont {M.~B.}\ \bibnamefont
  {Zvonarev}}, \bibinfo {author} {\bibfnamefont {V.~V.}\ \bibnamefont
  {Cheianov}},\ and\ \bibinfo {author} {\bibfnamefont {T.}~\bibnamefont
  {Giamarchi}},\ }\bibfield  {title} {\bibinfo {title} {{The time-dependent
  correlation function of the Jordan-Wigner operator as a Fredholm
  determinant}},\ }\href {https://doi.org/10.1088/1742-5468/2009/07/P07035}
  {\bibfield  {journal} {\bibinfo  {journal} {J. Stat. Mech.}\ }\textbf
  {\bibinfo {volume} {2009}},\ \bibinfo {pages} {P07035} (\bibinfo {year}
  {2009}{\natexlab{c}})},\ \Eprint {https://arxiv.org/abs/arXiv:0812.4059}
  {arXiv:0812.4059} \BibitemShut {NoStop}%
\bibitem [{Note1()}]{Note1}%
  \BibitemOpen
  \bibinfo {note} {For dynamic and finite temperature aspects of such a
  ``mobile impurity'' see also \cite
  {gamayun_impurity_Green_FTG_14,gamayun_impurity_Green_FTG_16,gamayun_protocol_impurity_18,10.21468/SciPostPhys.8.4.053,Gamayun2022}}\BibitemShut
  {NoStop}%
\bibitem [{\citenamefont {Gamayun}\ \emph {et~al.}(2015)\citenamefont
  {Gamayun}, \citenamefont {Pronko},\ and\ \citenamefont
  {Zvonarev}}]{gamayun_impurity_Green_FTG_14}%
  \BibitemOpen
  \bibfield  {author} {\bibinfo {author} {\bibfnamefont {O.}~\bibnamefont
  {Gamayun}}, \bibinfo {author} {\bibfnamefont {A.~G.}\ \bibnamefont
  {Pronko}},\ and\ \bibinfo {author} {\bibfnamefont {M.~B.}\ \bibnamefont
  {Zvonarev}},\ }\bibfield  {title} {\bibinfo {title} {{Impurity Green's
  function of a one-dimensional Fermi gas}},\ }\href
  {https://doi.org/10.1016/j.nuclphysb.2015.01.004} {\bibfield  {journal}
  {\bibinfo  {journal} {Nucl. Phys. B}\ }\textbf {\bibinfo {volume} {892}},\
  \bibinfo {pages} {83} (\bibinfo {year} {2015})},\ \Eprint
  {https://arxiv.org/abs/1410.1502} {1410.1502} \BibitemShut {NoStop}%
\bibitem [{\citenamefont {Gamayun}\ \emph {et~al.}(2016)\citenamefont
  {Gamayun}, \citenamefont {Pronko},\ and\ \citenamefont
  {Zvonarev}}]{gamayun_impurity_Green_FTG_16}%
  \BibitemOpen
  \bibfield  {author} {\bibinfo {author} {\bibfnamefont {O.}~\bibnamefont
  {Gamayun}}, \bibinfo {author} {\bibfnamefont {A.~G.}\ \bibnamefont
  {Pronko}},\ and\ \bibinfo {author} {\bibfnamefont {M.~B.}\ \bibnamefont
  {Zvonarev}},\ }\bibfield  {title} {\bibinfo {title} {{Time and
  temperature-dependent correlation function of an impurity in one-dimensional
  Fermi and Tonks-Girardeau gases as a Fredholm determinant}},\ }\href
  {https://doi.org/10.1088/1367-2630/18/4/045005} {\bibfield  {journal}
  {\bibinfo  {journal} {New J. Phys.}\ }\textbf {\bibinfo {volume} {18}},\
  \bibinfo {pages} {045005} (\bibinfo {year} {2016})},\ \Eprint
  {https://arxiv.org/abs/1608.08200} {1608.08200} \BibitemShut {NoStop}%
\bibitem [{\citenamefont {Gamayun}\ \emph {et~al.}(2018)\citenamefont
  {Gamayun}, \citenamefont {Lychkovskiy}, \citenamefont {Burovski},
  \citenamefont {Malcomson}, \citenamefont {Cheianov},\ and\ \citenamefont
  {Zvonarev}}]{gamayun_protocol_impurity_18}%
  \BibitemOpen
  \bibfield  {author} {\bibinfo {author} {\bibfnamefont {O.}~\bibnamefont
  {Gamayun}}, \bibinfo {author} {\bibfnamefont {O.}~\bibnamefont
  {Lychkovskiy}}, \bibinfo {author} {\bibfnamefont {E.}~\bibnamefont
  {Burovski}}, \bibinfo {author} {\bibfnamefont {M.}~\bibnamefont {Malcomson}},
  \bibinfo {author} {\bibfnamefont {V.~V.}\ \bibnamefont {Cheianov}},\ and\
  \bibinfo {author} {\bibfnamefont {M.~B.}\ \bibnamefont {Zvonarev}},\
  }\bibfield  {title} {\bibinfo {title} {Impact of the injection protocol on an
  impurity's stationary state},\ }\href
  {https://doi.org/10.1103/PhysRevLett.120.220605} {\bibfield  {journal}
  {\bibinfo  {journal} {Phys. Rev. Lett.}\ }\textbf {\bibinfo {volume} {120}},\
  \bibinfo {pages} {220605} (\bibinfo {year} {2018})},\ \Eprint
  {https://arxiv.org/abs/1402.6362} {1402.6362} \BibitemShut {NoStop}%
\bibitem [{\citenamefont {Gamayun}\ \emph {et~al.}(2020)\citenamefont
  {Gamayun}, \citenamefont {Lychkovskiy},\ and\ \citenamefont
  {Zvonarev}}]{10.21468/SciPostPhys.8.4.053}%
  \BibitemOpen
  \bibfield  {author} {\bibinfo {author} {\bibfnamefont {O.}~\bibnamefont
  {Gamayun}}, \bibinfo {author} {\bibfnamefont {O.}~\bibnamefont
  {Lychkovskiy}},\ and\ \bibinfo {author} {\bibfnamefont {M.~B.}\ \bibnamefont
  {Zvonarev}},\ }\bibfield  {title} {\bibinfo {title} {{Zero temperature
  momentum distribution of an impurity in a polaron state of one-dimensional
  Fermi and Tonks-Girardeau gases}},\ }\href
  {https://doi.org/10.21468/SciPostPhys.8.4.053} {\bibfield  {journal}
  {\bibinfo  {journal} {SciPost Phys.}\ }\textbf {\bibinfo {volume} {8}},\
  \bibinfo {pages} {053} (\bibinfo {year} {2020})}\BibitemShut {NoStop}%
\bibitem [{\citenamefont {Gamayun}\ \emph {et~al.}(2022)\citenamefont
  {Gamayun}, \citenamefont {Panfil},\ and\ \citenamefont
  {SantAna}}]{Gamayun2022}%
  \BibitemOpen
  \bibfield  {author} {\bibinfo {author} {\bibfnamefont {O.}~\bibnamefont
  {Gamayun}}, \bibinfo {author} {\bibfnamefont {M.}~\bibnamefont {Panfil}},\
  and\ \bibinfo {author} {\bibfnamefont {F.~T.}\ \bibnamefont {SantAna}},\
  }\bibfield  {title} {\bibinfo {title} {Mobile impurity in a one-dimensional
  gas at finite temperatures},\ }\bibfield  {journal} {\bibinfo  {journal}
  {Physical Review A}\ }\textbf {\bibinfo {volume} {106}},\ \href
  {https://doi.org/10.1103/physreva.106.023305} {10.1103/physreva.106.023305}
  (\bibinfo {year} {2022})\BibitemShut {NoStop}%
\end{thebibliography}%
\end{document}